\documentclass[a4paper]{article}
\usepackage{INTERSPEECH2020}
\usepackage{color}
\graphicspath{{fig/}}

\title{Attention-based Transducer for Online Speech Recognition}
\name{Bin Wang, Yan Yin, Hui Lin}
\address{LAIX Inc., Shanghai, China. }
\email{\{engine.wang, yin.yan, hui.lin\}@liulishuo.com}

\begin{document}
\newcommand{\vocab}{\mathcal{Y}}

\maketitle
\begin{abstract}
Recent studies reveal the potential of recurrent neural network transducer (RNN-T) for end-to-end (E2E) speech recognition. Among some most popular E2E systems including RNN-T, Attention Encoder-Decoder (AED), and Connectionist Temporal Classification (CTC), RNN-T has some clear advantages given that it supports streaming recognition and does not have frame-independency assumption. Although significant progresses have been made for RNN-T research, it is still facing performance challenges in terms of training speed and accuracy.  
We propose attention-based transducer with modification over RNN-T in two aspects. First, we introduce chunk-wise attention in the joint network. Second, self-attention is introduced in the encoder. Our proposed model outperforms RNN-T for both training speed and accuracy. 
For training, we achieves over 1.7x speedup.
With 500 hours LAIX non-native English training data, attention-based transducer yields $\sim$10.6\% WER reduction over baseline RNN-T. Trained with full set of over 10K hours data, our final system achieves $\sim$5.5\% WER reduction over that trained with the best Kaldi TDNN-f recipe.  After 8-bit weight quantization without WER degradation, RTF and latency drop to $0.34\sim 0.36$ and $268\sim 409$ milliseconds respectively on a single CPU core of a production server. 

\end{abstract}

\section{Introduction}
\label{sec:intro}

E2E models have recently become a popular choice for Automatic Speech Recognition (ASR), thanks to its simplicity of training process by integrating acoustic model (AM), pronunciation lexicon and language model (LM) components into one neural network component. E2E model directly transduces a sequence of input acoustic features to a sequence of output units (phonemes, characters, sub-words, words and etc.), which matches the inherent notion that ASR is a pure sequence to sequence task that maps a sequence of speech signals to a sequence of texts. 
Some widely used E2E approaches for sequence-to-sequence transduction are: 
a) Connectionist Temporal Classification (CTC) \cite{ctc1, ctc2},
b) Attention Encoder-Decoder (AED) \cite{bahdanau2014neural, chorowski2015attention, cho2014learning, bahdanau2016end}
and c) Recurrent Neural Network Transducer (RNN-T) \cite{rnnt, google_rnnt1}.
Among them AED becomes the most popular approach,
due to its training efficiency and promising results in a wide range of datasets \cite{LAS2, specaug}.
In order to apply AED in online speech recognition,
several online attention structures are proposed, such as chunk-wise attention \cite{sainath2018improving},
monotonic chunk-wise attention \cite{chiu2017monotonic}, triggered attention \cite{moritz2019triggered}.
However, applying such online attention mechanism usually causes performance degradation.
CTC is online streaming friendly. However, due to its frame-independency assumption, CTC requires an extra language model (LM) \cite{jasper} or combination with AED \cite{kim2017joint} for good WER results. 
RNN-T \cite{rnnt, google_rnnt1} is an elegant solution since it both supports streaming ASR and does not have frame-independency assumption. RNN-T has been successively used in mobile devices \cite{google_rnnt2} and production services \cite{li2019improving}. 
 
Most recent research for RNN-T focuses on exploring new model architectures, mainly for encoder and prediction network, and aiming at further improving the recognition accuracy.
\cite{tian2019self} proposes full self-attention transducer, in which self attention network is used for both encoder and decoder.
\cite{yeh2019transformer} introduces transformer transducer with causal convolution and truncated self-attention.
Latency controlled BLSTM  is used in encoder in \cite{rnnt_facebook}.
In \cite{sainath2019two},  LAS decoder is used to rescore the hypothesis generated by RNN-T. 
However, RNN-T is still facing several challenges. First, significant label imbalance between output units and blank exists in model training. Input sequence length is usually much larger than that of output sequence. For frame-by-frame alignment between input sequence and output sequence, large amount of blank symbols $\varnothing$ are inserted in the output sequence to denote ``null output'' \cite{rnnt}. This ends up with much more blank symbols than output units in the output sequence. Such unbalanced labels bias RNN-T towards predicting blank symbol, and generating more deletion errors. This causes bigger issue when words or word-pieces \cite{wordpiece} are used as output units, with which output sequence is even shorter. 
Second, as forward-backward algorithm is computational and memory expensive, 
it is hard to use a large mini-batch in RNN-T training \cite{li2019improving}.
As a result, the training speed of RNN-T is commonly much slower than AED \cite{LAS, LAS2} and CTC \cite{ctc1, ctc2}. 
Subsampling in encoder \cite{google_rnnt1}, to some extent, helps reduce the impact of label imbalance. But aggressively apply subsampling on input sequence to balance output units and blank usually causes performance degradation.  

We target on addressing above mentioned challenges in RNN-T by proposing attention-based transducer.
Our attention-based transducer is different from conventional RNN-T in two aspects.
First, we introduce chunk-wise attention in joint network, in which prediction network output attends on a chunk of encoder outputs. Output tokens are generated chunk by chunk. 
Second, we introduce self-attention in encoder to model contextual dependency. We also perform sub-sampling in encoder to reduce the time resolution of input sequence.
Regarding chunk-wise attention, the most relevant work  is in \cite{st}, 
where the authors propose a synchronous transformer by introducing chunk-wise attention and using forward-backward algorithm to optimize all alignment paths in training.
The major difference is, while \cite{st} aims to improve AED for online processing and chunk-wise attention is directly used to predict the output sequence, we focus on solving aforementioned RNN-T challenges and beam search is used to generate the output sequence.
Our attention-based transducer contributes in several aspects. 
First, adding chunk-wise attention helps reduce blank symbols in alignment, and effectively reduce the impact of label imbalance. Combining chunk-wise attention and encoder sub-sampling \cite{LAS} further helps alleviate label imbalance issue. 
Second, encoder and prediction network compose a grid of alignments, and forward-backward is conducted on the grid. This requires much more memory and computation than what is needed in AED or CTC. Introducing chunk-wise attention significantly reduce the grid size, and as a result, memory consumption and computation. Such improvement allows us to use larger mini-batch in training and helps speedup training significantly. 
Also adding self-attention \cite{selfatt}  in encoder better models the contextual dependency, thus helps improve recognition accuracy. 

To evaluate the performance of our attention-based transducer for online speech recognition, 
we train our model with LAIX L2 English dataset spoken by Chinese,
and compare with LAS, conventional RNN-T, and Kaldi system in terms of word error rate (WER), real time factor (RTF) and latency.
With 500-hour training data, 
our attention-based transducer achieves about 14.4\% relative WER reduction over a LAS model with unidirectional encoder,
and 10.6\% relative WER reduction over a baseline RNN-T model.
For our full 10K-hour L2 English training set, 
our attention-based transducer achieves $\sim$5.5\% relative WER reduction over TDNN-f model trained with the most updated Kaldi recipe.
With 8-bit weight quantization and almost no accuracy degradation, RTF and latency drops to about $0.34\sim 0.36$ and 268$\sim$409 milliseconds respectively on a single CPU core of a production server.

The rest of the paper is organized as follows.
In Session 2, we review the recurrent neural network transducer (RNN-T).
Then attention-based transducer is introduced in Session 3.
We evaluate attention-based transducer with 500h and 10kh LAIX English tasks spoken by Chinese in Session 5.
Finally we conclude our study in Session 6.

\section{Recurrent neural network transducer}
\label{sec:rnnt}

\begin{figure}
	\includegraphics[width=0.95\linewidth]{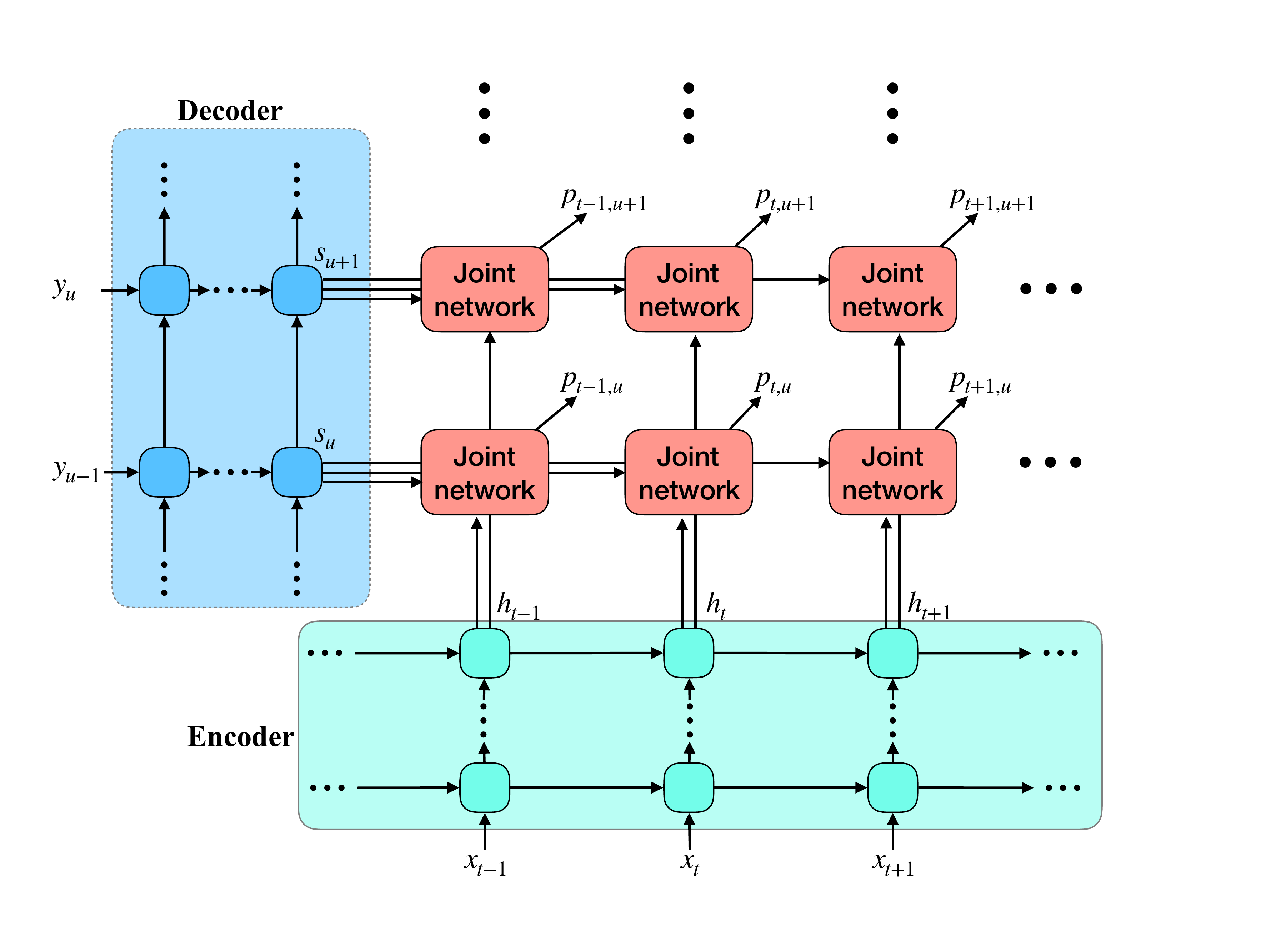}
	\vspace{-10pt}
	\caption{recurrent neural network transducer (RNN-T)}
	\label{fig:old-rnnt}
	\vspace{-15pt}
\end{figure}

Recurrent neural network transducer (RNN-T) is first proposed in \cite{rnnt}
to directly transform an input sequence to an output sequence, 
without knowing the alignment information.
Denote by $x = (x_1, \cdots, x_T)$ an input sequence of length $T$,
and by $y=(y_1, \cdots, y_U)$ an output sequence of length $U$,
where $y_u \in \vocab$ for $1 \leq u \leq U$ and $\vocab$ is the set of output units. 
There are three major components in a RNN-T model (shown in figure \ref{fig:old-rnnt}).
\begin{enumerate}
	\item \textbf{Encoder:}  Encoder can be thought of an acoustic model, which processes an input sequence $x$ and generates deep hierarchical features $h=(h_1, \cdots, h_T)$. 
	Commonly used encoder architectures include LSTM \cite{google_rnnt1, google_rnnt2},  latency controlled BLSTM \cite{rnnt_facebook}, deep Convolutional Neural Network (CNNs) \cite{jasper} and self-attention network \cite{pham2019very}. 
	\item \textbf{Decoder:}  Decoder (or called ``prediction network'' in \cite{rnnt}) is analogous to language model, which takes previous unit $y_{u-1}$ and the hidden states as inputs, and outputs vector $s_{u}$ as representation of all history units $y_1, \cdots, y_{u-1}$. 
	
	\item \textbf{Joint network:} Encoder output $h_t$ and decoder output $s_u$ are fed into joint network to compute output logit $z_{t, u}$ for each time step $t$ and output position $u$.
	The output logits $z_{t,u}$  are then passed to the softmax layer and distribution $p_{t,u}$ over the set of output units and blank, denoted as $\vocab \cup \{\varnothing\}$, is generated.	 
\end{enumerate}

The output distribution $p_{t,u}$ ($1 \leq t \leq T$, $1 \leq u \leq U$) defines a two-dimensional grid, where a path from $(t=1,u=1)$ to $(t=T, u=U)$ denotes an alignment between an input sequence $x$ and an output sequence $y$, and the probability of the path is computed according to $p_{t,u}$.
Forward-backward algorithm is used to compute the transduction probability $p(y|x)$, by summing up probabilities of all alignment paths.
RNN-T is trained to maximize the logarithmic likelihood (i.e. $\log p(y|x)$), and back-propagation is used to update the parameters.
 
Though RNN-T supports streaming ASR, it is still facing several challenges.
First, it suffers from label imbalance issue.
In each alignment path, $T$ blank symbols are inserted into the output sequence of length $U$. 
In speech recognition, input sequence length $T$ (i.e. the acoustic feature length) is usually much larger than output sequence length $U$, especially in case where sub-word or word are used as output unit. Blank symbol usually dominates a label sequence, 
and probability of blank symbol become dominant during model training.
As a result, model is biased toward generating blank symbol. 
In our experiments, the number of deletion errors is about 8 times larger than the number of insertion errors, leading to a much degraded recognition accuracy. 

Second, compared with AED \cite{LAS},
training RNN-T is computational and memory expensive.
The output of joint network is a 3-dimensional matrix of shape $T \times U \times (|\vocab| + 1)$, where $T$ is the length of encoder output, $U$ is the length of output sequence, and $|\vocab|$ is the number of output units.
Large memory consumption with such 3-dimension matrix has big impact on mini-batch size in model training.
As a result, RNN-T training is usually much slower than those of other sequence-to-sequence approaches, such as AED and CTC. 
In our experiments, batch size in RNN-T training is 20,000 frames,
which is only a quarter of that used  in AED training with same encoder and decoder sizes.

Finally, joint network only uses encoder output at current time step $h_t$, without considering the contextual information, i.e. $(\cdots, h_{t-1}, h_{t+1}, \cdots)$. 
It has been shown not only in conventional hybrid system\cite{tdnn}, but in end-to-end system \cite{das2019advancing} that,
contextual information, especially the look-ahead information, is helpful to improve model accuracy.

\section{Attention-based Transducer}
\label{sec:mymodel}

\begin{figure}
	\includegraphics[width=0.99\linewidth]{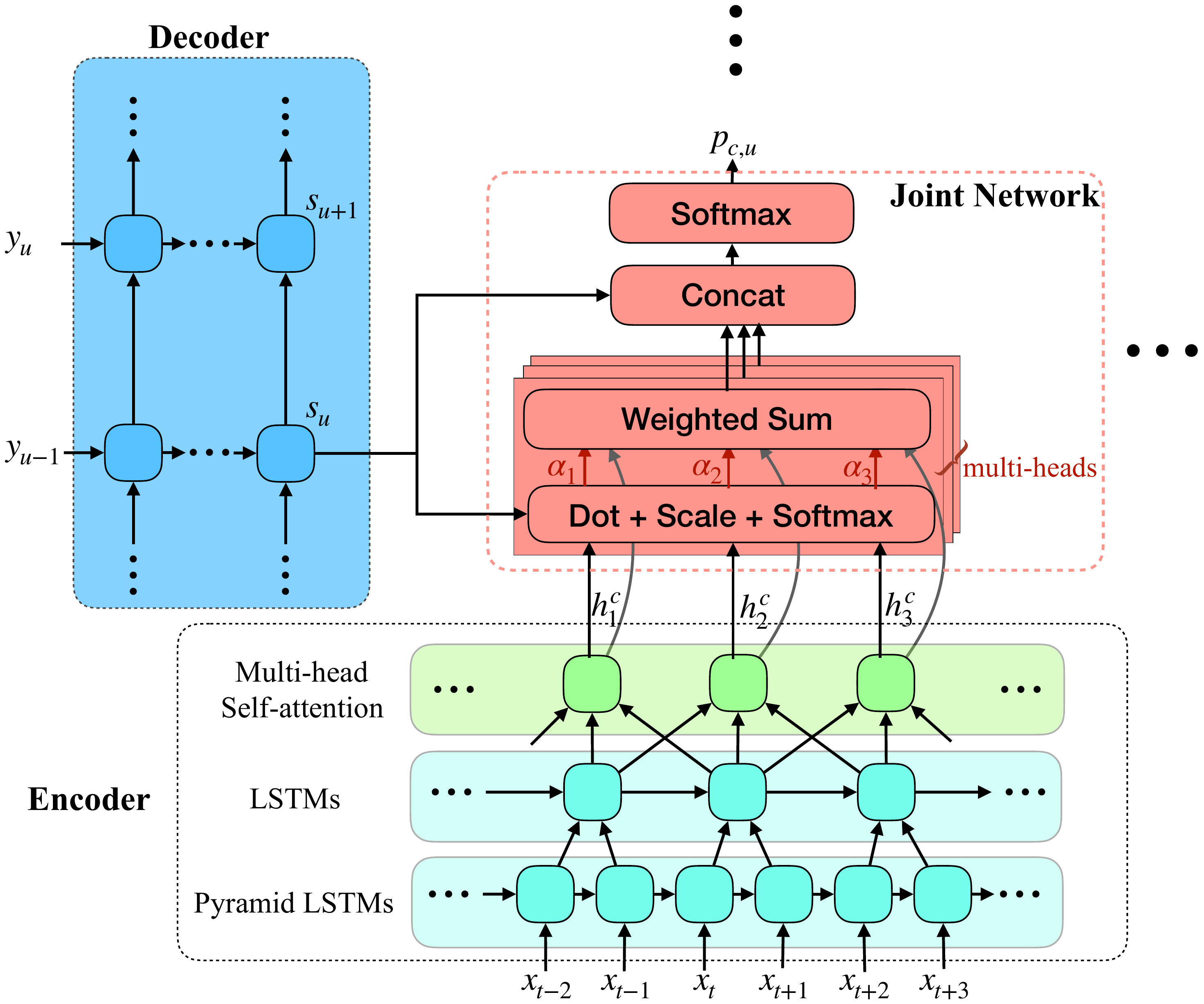}
	\vspace{-10pt}
	\caption{Model structure of attention-based transducer}
	\label{fig:new-rnnt}
	\vspace{-15pt}
\end{figure}

In this session, we propose an attention-based transducer, 
by incorporating the attention mechanism into the transducer framework.
The model architecture is shown in figure \ref{fig:new-rnnt} and the details are described as follows.

\textbf{Encoder:} An input sequence $x$ is fed into unidirectional pyramid Long Short Term Memory (pLSTM) \cite{lstm, LAS} layers. 
In each layer, the adjacent 2 outputs of LSTM are concatenated and then fed into the next LSTM layer, reducing the length of its input sequence by a factor of  2.
Given $n_p$ as the number of pLSTM layers,
the length of final encoder output sequence is reduced by a factor of $\mu = 2n_p$.
Such sub-sampling approach helps speed up training and inference of RNN-T \cite{google_rnnt2}.
Then several LSTM layers are stacked on top,
and followed by a local multi-head self-attention \cite{selfatt, das2019advancing} layer, to learn contextual dependency.
Assume the output of the last LSTM layer is $h^{lstm}_t \in R^{d}$ ($1 \leq t \leq \frac{T}{\mu}$),
where $\mu$ is the total down-sampling factor of pLSTM layers,
the output of self-attention at time step $t$ is computed as follows:
\begin{align}
	s_i &= \frac{(Qh^{lstm}_t)^T Kh^{lstm}_i}{\sqrt{d/n_{att}}}, \quad i = t-\tau, \ldots, t+\tau \\
	\alpha_i &= \frac{\exp(s_i)}{\sum_{j=t-\tau}^{t+\tau} \exp(s_j)}, \quad  i = t-\tau, \ldots, t+\tau \\
	c_t &= \sum_{i=t-\tau}^{t+\tau} \alpha_i Vh^{lstm}_i \\
	h_t &= \text{LayerNorm}(c_t + h^{lstm}_t)
\end{align}
where $\tau$ is the context length,
$n_{att}$ is the number of attention heads,
$Q, K, V \in R^{\frac{d}{n_{att}} \times d}$ are parameters, 
$\text{LayerNorm}$ denotes layer normalization \cite{layernorm}.
Finally, the outputs of all attention heads are concatenated.

\textbf{Decoder:} 
Decoder consists of  2 LSTM layers, 
which is commonly used in conventional RNN-T.
To alleviate the discrepancy between training and inference,
at each step of training, we either feed in the true label or a random label sampled from the vocabulary $\vocab$, and random label is sampled based on a probability $p_{ss}$.
This strategy is similar to scheduled sampling proposed in \cite{scheduled_sampling}, but based on different distributions. 

\textbf{Joint network:}
The biggest difference between attention-based transducer and conventional RNN-T is in the joint network component.
In conventional RNN-T, the joint network only accepts the output of encoder at current time step $h_t$.
In our attention-based transducer, joint network takes a chunk of encoder outputs as input , and has decoder output attending on the chunk of encoder outputs.
First, the outputs of the encoder are split into fix-length chunks without overlapping.
Denoting $h^c_i \in R^{d}$ ($1 \leq c \leq \frac{T}{\mu w}$, $1 \leq i \leq w$)  as the $i$-th vector of $c$-th chunk, where $w$ is the chunk width,
the attention weights are computed as follows:
\begin{align}
	a^{c, u}_i &= \frac{(\hat{Q}s_u)^T \hat{K} h^c_i}{\sqrt{d/n_{att}}}, \quad i = 1, \ldots, w \\
	\alpha^{c, u}_i &= \frac{\exp(a^{c,u}_i)}{\sum_{j=1}^{w} \exp(a^{c, u}_j)}, \quad i = 1, \ldots, w \\
\end{align}
where $\hat{Q}, \hat{K} \in R^{\frac{d}{n_{att}} \times d}$ are the parameter matrices.
After weighted summation, we get attention output,
\begin{equation}
	o_{c,u} = \sum_{i=1}^{w} \alpha^{c,u}_i \hat{V}h^c_i
\end{equation}
where $\hat{V} \in R^{\frac{d}{n_{att}} \times d}$ is the parameter matrix.
Similar to self-attention in encoder,
multi-head attention with head number $n_{att}$ is used.
Then outputs of $n_{att}$ attentions and  decoder $s_u$ are concatenated and passed to a Softmax layer to get distribution $p_{c, u}$ over output units and blank, i.e. $\vocab \cup \{\varnothing\}$.

Similar to conventional RNN-T, forward-backward algorithm is used in model training.
During inference, beam search \cite{rnnt} is used to generate the best hypothesis sequence chunk by chunk.
To simplify the algorithm and make it ``less computationally intensive'', we use the version proposed in \cite{google_rnnt1} by skipping the summation  over prefixes in $pref(y)$ (See Algorithm 1 in \cite{rnnt}).

Attention-based transducer resolves the earlier mentioned challenges with conventional RNN-T.
First, length of encoder output sequence is reduced by a factor of $\mu w$, where $\mu$ is the total down-sampling factor of encoder and $w$ is the chunk width of chunk-wise attention in joint network.
As a result, the number of blank symbols in each alignment are significantly reduced from $T$ to $\frac{T}{\mu w}$. This alleviates the training label imbalance between blank and output units. 
Second, the output of joint network $p_{c,u}$ ($1 \leq c \leq \frac{T}{\mu w}$, $1 \leq u \leq U$) is a matrix of shape $\frac{T}{\mu w} \times U \times (|\vocab|-1)$, 
which is $\mu w$ times less than the output matrix of conventional RNN-T (i.e. $T \times U \times (|\vocab|-1)$).
This allows us to train model with larger mini-batch. 
In our experiments, the batch size of attention-based transducer is 2 time larger than conventional RNN-T.
Moreover, multi-head self attention helps model better contextual dependency, thus improve recognition accuracy.

\section{Experiments}
\label{sec:exps}

We conduct experiments to evaluate how attention-based transducer performs on our in-house data of non-native English, spoken by second language (L2) earners whose native language is Chinese. The data is mostly collected from LAIX language learning App. The full training dataset of more than 10k hours used in this study covers,

\begin{enumerate}
	\item \textbf{Read speech} (5670 hours): this contains mostly read speech of L2 learners reading given sentences, and also a small part of native speaker's read speech.
	\item \textbf{Spontaneous speech} (6550 hours): this contains speech of L2 learners answering open and semi-open questions, and L2 learner's conversational speech.  
	
\end{enumerate}

Our first experiment is based on 500-hour training subset of conversational speech. In this experiment, we compare our proposed attention-based transducer with conventional RNN-T and AED.  
In the other experiment, the full 10kh training set is used to build our best RNN-T system based on the best configuration tuned from 500-hour experiment, and we compare our best system with a strong Kaldi's hybrid ASR production system trained with same data. 
for both experiments, a separate set of conversational speech ($\sim$30 hours) is used as evaluation data.

\subsection{Experiment on 500 hours of speech}

\begin{table}
\centering
\begin{tabular}{l|ccc}
	\hline
	Models & WER & RTF & Latency\\
	\hline \hline
	LAS & 18.67  & - & -  \\
	baseline RNN-T & 17.90 & 0.36 & 404 ms \\
	\hline
	att. transducer $\tau=0, w=4$ & 18.77 & 0.33 & 97 ms \\
	att. transducer $\tau=1, w=4$ & 17.13 & 0.33 & 159 ms\\
	att. transducer $\tau=2, w=4$ & 16.05 & 0.33 & 250 ms \\
	att. transducer $\tau=4, w=4$ & \textbf{15.98} & 0.34 & 409 ms \\
	att. transducer $\tau=4, w=8$ & 16.18 & 0.31 & 402 ms \\
	att. transducer $\tau=4, w=16$ & 16.76 & 0.31 & 411 ms \\
	\hline
\end{tabular}
	\caption{Results on 500 hours of Liulishuo L2 English speech.}
	\label{tab:res1}
	\vspace{-20pt}
\end{table}

\begin{table}
\centering
\begin{tabular}{l|ccc}
	\hline
	Models & Ins. & Del. & Sub. \\
	\hline
	baseline RNN-T & 127 & 1046 & 1585 \\
	\hline
	att. transducer $\tau=4, w=4$ & 171 & 715 & 1575 \\
	\hline
\end{tabular}
	\caption{Insertion errors (Ins.), deletion errors (Del.) and substitution errors (Sub.) after recognizing the test speech.}
	\label{tab:errs}
	\vspace{-20pt}
\end{table}

The encoder of our attention-based transducer consists of 3 unidirectional pLSTM layers, followed by 2 LSTM layers. An input sequence is down-sampled by a factor of $\mu=8$ after 3 pLSTM layers.
All LSTM layers (including pLSTM) have 1024 hidden units (i.e. $d=1024$).
The decoder consists of 2 LSTM layers each with 512 hidden units.
4-head attention ($n_{att}=4$) is used in both encoder and joint network.
Context length $\tau$  and chunk width $w$ are fine-tuned in the experiment.
For fair comparison, the same encoder and decoder architecture are used in baseline RNN-T.

The input $x$ is a sequence of 80-dimensional log filterbank energies with frame length 25 milliseconds and frame step 10 milliseconds.
The output units contain 500 word-pieces generated by byte-pair encoding  \cite{wp}.
During training, a LAS model \cite{LAS} is used to pre-train the encoder.
In our experiment, this significantly improves RNN-T performance.
TensorFlow \cite{tensorflow2015-whitepaper} is used for our experiment.
We further apply model quantization to reduce the model size and also speed up model inference. TFLite is used to quantize weight parameters to 8-bit integer. 
Our recognition engine processes input speech stream chunk-by-chunk, each with 100 milliseconds.  
Beam search with beam size 8 is used to generate the best hypothesis.

To evaluate our proposed model for streaming ASR, we report word error rate (WER), real time factor (RTF) and latency in our experiments. 
For RTF we take the average over all chunks. 
Latency represents the total response time, which is the averaged recognition time of each speech chunk and its look ahead size (context length $\tau$).  
Intuitively, latency should be approximately equal to the summation of the recognition time of a chunk and the look ahead time ($\tau \times 8 \times 10$ ms where $8$ is is the total down-sampling factor of encoder, $10$ ms is the frame step).
RTF and latency are measured on a single CPU core (Intel(R) Xeon(R) Platinum 8269CY CPU @ 2.50GHz) of a production server. 

As we can see from the results shown in Table \ref{tab:res1},
attention-based transducer (``att. transducer'') with $\tau=4, w=4$ achieves the lowest WER 15.98,  
with 14.4\% relative WER reduction over LAS model,
10.7\% relative WER reduction over baseline RNN-T model.
From Figure \ref{tab:errs}, our attention-based transducer significantly alleviate label imbalance issue. Deletion errors are reduced by about 30\% compared with baseline RNN-T model. 
Increasing context length significantly reduces WER but with much increased latency. 

\subsection{Experiment on 10 thousand hours of speech}
 
\begin{table}
\centering
\begin{tabular}{l|ccc}
    \hline
	Models & WER & RTF & Latency\\
	\hline \hline
	hybrid ASR  & 10.9 & 0.15 & 385 ms \\
	\hline
	att. transducer $\tau=2, w=4$ & 10.3 & 0.19 & 300 ms  \\
	\hline
\end{tabular}	
	\caption{Results on 10 thousands hours of Liulishuo L2 English speech}
	\label{tab:res2}
	\vspace{-20pt}
\end{table}
 
Our final attention-based transducer system is built with our full training set of size over 10k hours using the configuration we tuned in 500-hour task. 
Our final system is compared with Kaldi's hybrid system.
For Kaldi's hybrid system, TDNN-f contains 17 sub-sampled time-delay neural network layers with low-rank matrix factorization (TDNNF) \cite{povey2018semi}, and trained based on the lattice-free MMI \cite{povey2016purely} recipe in Kaldi toolkit \cite{kaldi}.
Language model is a 3-gram language model (LM) with modified Kneser-Ney smoothing trained using the SRILM toolkit \cite{SRILM}.

The results are shown in Table \ref{tab:res2}.
Our final system outperforms Kaldi's hybrid system with relative WER reduction 5.5\%, and achieves latency about 300 ms with a small contextual width $\tau=2$. Though our final transducer system has similar RTF as Kaldi's hybrid system, it achieves lower latency mainly due to the use of small look ahead size. The system is being launched in LAIX's AI-teacher product, and we plan to launch an on-device transducer system soon. 

\section{Conclusion}

In this paper, we propose attention-based transducer to address the challenges in conventional RNN-T model. We introduce chunk-wise attention in joint network. Chunk-wise attention helps alleviate label imbalance between sub-word units and blank. it also reduces the size of alignment grid, thus requires much less memory in training. This allows us to use bigger batch size, and achieve significant training speedup. We also introduce self attention layer in encoder, which models better contextual dependency and yields improved recognition accuracy with acceptable latency. 
From another perspective, we can also view attention-based transducer as an approach to merge the ideas behind AED and RNN-T into the same framework. 
Attention-based transducer with chunk size $w=1$ is conventional RNN-T. When $w$ is equal to the length of encoder output, attention-based transducer is quite similar to AED.
Our experiments show superior performance of the proposed attention-based transducer over conventional RNN-T and Kaldi's TDNN-f models. The system is being launched in LAIX’s AI-teacher product. 
Future work will include exploring adaptation strategies to improve model generalization capability to new domains, and also further optimizing model size and computation cost for on-device application.

\bibliographystyle{IEEEtran}
\bibliography{mybib}

\end{document}